\documentclass[aps,prd,11pt,superscriptaddress,preprintnumbers,
amsfonts,amssymb,amsmath]{revtex4}

\begin{document}

\title{``Detuned'' $ \boldsymbol{f(R)} $ gravity and dark energy}

\author{Nathalie Deruelle}
\affiliation{
APC, (CNRS, Universit\'e Paris 7, CEA, Observatoire de Paris), Paris, France
}
\affiliation{
Racah Institute of Physics, Hebrew University, Jerusalem, Israel
}
\author{Misao Sasaki}
\affiliation{
Yukawa Institute for Theoretical Physics, Kyoto University, Kyoto, Japan
}
\author{Yuuiti Sendouda}
\affiliation{
Yukawa Institute for Theoretical Physics, Kyoto University, Kyoto, Japan
}
\date{April 2, 2008}

\begin{abstract}
In gravity theories derived from a $f(R)$ Lagrangian, matter is usually supposed to be minimally coupled to the metric, which hence defines a ``Jordan frame.''
However, since the field equations are fourth order, gravity possesses an extra degree of freedom on top of the standard graviton, as is manifest from its equivalent description in the conformally related, Einstein, frame.
We introduce explicitly this extra scalar degree of freedom in the action and couple it to matter, so that the original metric no longer defines a Jordan frame.
This ``detuning'' puts $f(R)$ gravity into a wider class of scalar--tensor theories.
We argue that a ``chameleon-like'' detuning tracing the background matter density may provide purely gravitational models which account for the present acceleration of the universe and evade local gravity constraints.
\end{abstract}

\preprint{YITP-08-16}

\maketitle

\section{Introduction} 
Observations have shown that the observable universe is well described by a Friedmann--Lema{\^\i}tre cosmological model whose scale factor is presently accelerating \cite{accel}.
The simplest way to account for this acceleration is to add a cosmological constant to Einstein's equations, the challenge being then to explain why it is so small \cite{cosmoconst} and starts to dominate now.
An even more economic way would be to show that this acceleration is an artefact of the averaging process, but no such satisfactory model has yet been proposed (see e.g.\ \cite{Buchert} for a review).
As for the models giving a dynamical origin to this ``dark energy'' they fall into two broad categories: the ``quintessence'' models \cite{quint} where the acceleration is due to a scalar field coupled to gravity, and the ``modified gravity'' models \cite{modgrav} where it is due to a modification of the Einstein field equations.
These two classes of models are in fact related as most models fall into the scope of scalar--tensor theories of gravity \cite{scaten}.

We shall here concentrate on dark energy theories where the Lagrangian for gravity is taken to be a function $f(R)$ of the Ricci scalar $R$ \cite{modgrav}.
As is well known \cite{modgrav,TeyTou}, gravity is mediated in such theories by the usual graviton plus a scalar, dubbed ``scalaron'' in \cite{Staron}.
However, in most models (see e.g.\ \cite{Cognola} for a review and recent results) matter is supposed to be minimally coupled to the graviton, that is, the metric, only.
Following our suggestion in \cite{DerSasSen}, we propose here to also couple it to the scalaron, in a conformal way, so that there still exists a ``Jordan frame,'' different from the original one, where matter is minimally coupled to a metric (see \cite{gravcouple,Tsujikawa} for some special examples of such non-minimal couplings).
We regard this Jordan frame in which matter is minimally coupled to gravity as the physical frame.
As we shall argue, a ``chameleon-like detuning,'' tracing the background matter density, may provide a way to alleviate some problems faced by the minimally coupled models proposed in the literature.

\section{Detuned $ \boldsymbol{f(R)} $ gravity}
Consider the following action for gravity in the presence of matter,
\begin{equation}
S[g_{ij};s]
=\frac{1}{2\kappa}\int_{\cal M}\mathrm d^4x\sqrt{-g}[f'(s)R-(s f'(s)-f(s))]
+S_{\rm m}[\Psi;\tilde g_{ij}=\mathrm e^{2C(s)}g_{ij}]\,.
\label{eqno1}
\end{equation}
Here $\mathrm ds^2=g_{ij}\mathrm dx^i\mathrm dx^j$ is the line element in ${\cal M}$ with the coordinates $x^i$ and metric $g_{ij}(x^i)$, signature $-+++$ and determinant $g$;
$R=g^{ij}R_{ij}=g^{ij}\partial_k\Gamma^k_{ij}-\cdots$ is the scalar curvature, $\kappa$ is a coupling constant;
the functions $f(s)$ and $C(s)$ of the scalaron field $s(x^i)$ are for the moment arbitrary (we suppose $f'(s)>0$); $\Psi(x^i)$ denotes the matter fields and $S_{\rm m}$ their action.
The novelty (as far as we are aware in this context) is that, as suggested in \cite{DerSasSen}, we allow for an a priori arbitrary (conformal) coupling of matter to the scalaron.
The field equations stemming from (\ref{eqno1}) are
\begin{equation}
\begin{aligned}
& f''(s)(s-R)=2\kappa\,C'(s)T\,,\\
& f'(s)G_{ij}+{1\over2}g_{ij}(sf'(s)-f(s))
+g_{ij}\square f'-D_{ij}f'=\kappa T_{ij}\,,
\end{aligned}
\label{eqno2}
\end{equation}
where a prime denotes derivation with respect to $s$;
$G_{ij}=R_{ij}-{1\over2}g_{ij}R$ is the Einstein tensor;
$T^{ij}={2\over\sqrt{-g}}{\delta S_{\rm m} \over\delta g_{ij}}$ is the stress--energy tensor of matter (for example, $T_{ij}=\mathrm e^{4C(s)}[(\tilde\rho+\tilde p)u_iu_j+\tilde p g_{ij}]$ for a perfect fluid, with $g^{ij}u_iu_j=-1$ and $\tilde\rho$ and $\tilde p$ representing its energy density and pressure as measured in a locally inertial frame), and $T=g_{ij}T^{ij}$.
We note that $T^{ij}$ is not conserved: $D_jT^j{}_i=T\,C'(s)\partial_is$.

In the absence of coupling to the scalaron ($C(s)=0$), we have that
matter is minimally coupled to $g_{ij}$ ($D_jT^j{}_i=0$), so that ${\cal M}$
is the ``Jordan frame'' representing physical spacetime, and we have
$s=R$ (we suppose $f''\neq0$).
Thus the equations reduce to those obtained by varying the action ${1\over2\kappa}\int_{\cal {M}}\mathrm d^4x\sqrt{-g}f(R)+S_{\rm {m}}$ with respect to the metric alone.
In this case, as has been shown in \cite{modgrav},
simple models such that $f(R)\propto 1/R^n$ for small $R$ 
may account for the present acceleration of the universe.

In presence of coupling, gravity is ``detuned'' in that ${\cal M}$ 
 is no longer the Jordan frame.
However (\ref{eqno1}) can be rewritten as 
(we follow the notation of \cite{Esposito})
\begin{equation}
\tilde S[\tilde g_{ij}; \Phi]
={1\over2\kappa}\int_{\tilde{\cal M}}\mathrm d^4x\sqrt{-\tilde g}
\left[\Phi\tilde R-{\omega(\Phi)\over\Phi}(\tilde\partial\Phi)^2
-2U(\Phi)\right]+S_{\rm m}[\Psi;\tilde g_{ij}]\,,
\label{eqno3}
\end{equation}
where $\tilde S$ and $S$ are equal up to a boundary term;
$\mathrm d\tilde s^2=\tilde g_{ij}\mathrm dx^i\mathrm dx^j$ is the line element in the Jordan frame $\tilde{{\cal M}}$ with metric $\tilde g_{ij}=\mathrm e^{2C(s)}g_{ij}$;  
the potential $U(\Phi)$ and the Brans--Dicke function $\omega(\Phi)$ are given by ($s$ being now a parameter)
\begin{equation}
\begin{aligned}
&\Phi(s)=f'(s)\mathrm e^{-2C(s)}\,,\quad
U(s)={sf'(s)-f(s)\over2}\mathrm e^{-4C(s)}\,,\\
&\omega(s)=-{3K(s)(K(s)-2)\over 2(K(s)-1)^2}\quad
\hbox{with}\quad
K(s)\equiv{\mathrm dC\over\mathrm d(\ln\sqrt{f'})}\,.
\end{aligned}
\label{eqno4}
\end{equation}
When $C(s)=0$, we recover scalar--tensor theories 
with a vanishing Brans--Dicke function, $\omega=0$.
This property rules out models such as $f(R)=R-\mu^{2(n+1)}/R^n$ \cite{modgrav}, as solar system tests impose $\omega>40,000$ \cite{Bertotti:2003rm} when the potential can be ignored on local scales, see \cite{scaten,Chiba}.
We also note that in regimes where $K(s)\to\infty$, we have $\omega(s)\to-3/2$ and the theory resembles  Palatini-$f(R)$ gravity with matter coupled to the metric only \cite{Vollick}.
From the expression (\ref{eqno4}) of the Brans--Dicke function one may see what we are aiming at: a coupling function $K(s)$ which vanishes on cosmological scales in order to account for the present-day acceleration of the universe, but which tends to $1$ on local gravity scales in order to comply with the solar system constraints.
 
Finally it will be convenient to work in the ``Einstein frame'' where the action (\ref{eqno1}) becomes
\begin{equation}
S^*[g_{ij}^*;\varphi]
={2\over\kappa}\int_{{\cal M}^*}\mathrm d^4x\sqrt{-g^*}
\left[{1\over4}R^*-{1\over2}(\partial^*\varphi)^2-V(\varphi)\right]
+S_{\rm m}[\Psi;\tilde g_{ij}=\mathrm e^{2C(\varphi)}f'(\varphi)^{-1}g_{ij}^*]\,,
\label{eqno5}
\end{equation}
where, again, $S^*$ differs from $S$ by a boundary term;
 $\mathrm ds^{*2}=g^*_{ij}\mathrm dx^i\mathrm dx^j$ is the line element in the Einstein
 frame ${\cal M}^*$ with metric $g_{ij}^*$; $C(\varphi)$ and $f'(\varphi)$
 stand for $C(s(\varphi))$ and $f'(s(\varphi))$;
 the scalar field $\varphi$ and the potential $V(\varphi)\equiv V(s(\varphi))$
 are given under a parametric form as
\begin{equation}
\varphi(s)=\sqrt{3}\ln\sqrt{f'(s)}\,,\quad
V(s)={sf'(s)-f(s)\over4f'^2(s)}\,.
\label{eqno6}
\end{equation}
In this Einstein frame the field equations (\ref{eqno2}) read
\begin{align}
&\square^*\varphi-{\mathrm dV\over\mathrm d\varphi}
={1-K(\varphi)\over2\sqrt{3}}\kappa T^*\,,
\label{KGeq}\\
&G_{ij}^*-2\partial_i\varphi\partial_j\varphi
+g_{ij}^*[(\partial^*\varphi)^2+2V(\varphi)]=\kappa T^*_{ij}\,.
\label{Eeq}
\end{align}
The stress--energy tensor, 
$T^{*ij}={2\over\sqrt{-g^*}}{\delta S_{\rm m}\over\delta g^*_{ij}}$
 (for example, $T^*_{ij}=\mathrm e^{4C(\varphi)} f'(\varphi)^{-2}
[(\tilde\rho+\tilde p)u^*_iu^*_j+\tilde p g^*_{ij}]$
 for a perfect fluid, with $g^{*ij}u^*_i u^*_j=-1$), 
is not conserved: $D^*_jT^{*j}{}_i=-[(1-K(\varphi))/\sqrt 3]T^*\partial_i\varphi$.
The function $K(\varphi)\equiv K(s(\varphi))$ is defined by (\ref{eqno4}) and (\ref{eqno6}).
When $K=0$ Eqs.~(\ref{eqno6}--\ref{Eeq}) are the Einstein frame version of the standard, minimally coupled $f(R)$ gravity.
In the ``maximally detuned'' case $K=1$ (advocated in e.g.\ \cite{maxdetune}) the Jordan and Einstein frames coincide, and the theory reduces to General Relativity minimally coupled to a scalar field with a potential given by (\ref{eqno6}).
For example, if $f(R)\propto 1/R^n$, $V(\varphi)$ decays exponentially and the theory is akin to the scaling field models of dark energy first studied in \cite{Ferreira}.
Finally, for $C(s)\propto\ln f'(s)$, that is, $K=\mathrm{constant}$ (see \cite{Tsujikawa}), the theory resembles the non-minimally coupled quintessence models \cite{mcquint}, including the ``chameleon'' models \cite{chamel}, albeit with a potential which is typically exponential for large $\varphi$ and hence different from the fiducial quintessential model \cite{fquint}: $V(\varphi)=M^4\mathrm e^{M^n/\varphi^n}$.

\section{Cosmological evolution}
When the metric is taken to be that of a spatially flat Robertson--Walker spacetime, $\mathrm ds^{*2}=-\mathrm dt^2+a^{*2}(t)\mathrm d\vec{x}^2$, and matter to be pure dust, the Einstein frame field equations (\ref{KGeq}--\ref{Eeq})
reduce to
\begin{equation}
3H^{*2}-\dot\varphi^2-2V(\varphi) =\kappa\rho^*\,,\quad
\ddot\varphi+3H^*\dot\varphi+{\mathrm dV\over\mathrm d\varphi}
={1-K(\varphi)\over2\sqrt{3}}\kappa\rho^*\quad
\hbox{with}\quad
\rho^*={\rho^*_0\,\mathrm e^{C(\varphi)-{\varphi\over\sqrt{3}}}\over a^{*3}}\,,
\label{eqno8}
\end{equation}
where $H^*={\dot a^*/a^*}$ and a dot representing derivative with respect to $t$.
The observable Jordan frame scale factor is given by $\tilde a(\tilde t)=a^*\mathrm e^{C(\varphi)-{\varphi/\sqrt{3}}}$ where $\tilde t$ is the cosmic time in the Jordan frame ($\mathrm d\tilde t=\mathrm e^{C(\varphi)-{\varphi/\sqrt{3}}}\mathrm dt$).

Suppose that, as in \cite{modgrav}, $f(s)\approx -1/(\ell^2\bar s^n)$ 
($n>0$) for small $\bar s$, where $\bar s\equiv\ell^2s$ with $\ell$ being
of the order of the Hubble radius today.
Then $\varphi$ is large and $V(\varphi)\propto\mathrm e^{-\lambda\varphi}$ with 
$\lambda={2\over\sqrt{3}}(n+2)/(n+1)$.
When matter has become negligible, the Einstein frame scale factor and the scalar field do not depend on $C(\varphi)$, and they are given by $a^*\propto t^q$ with $q=3(n+1)^2/(n+2)^2$ and $\varphi\sim p\ln t$ with $p=\sqrt{3}(n+1)/(n+2)$ \cite{quint,modgrav,Ferreira}.
On the other hand the Jordan frame scale factor $\tilde a(\tilde t)$ does 
depend on the coupling function $C(\varphi)$.
If $C(\varphi)\to0$ for large $\varphi$, then 
we have $\tilde a(\tilde t)\propto \tilde t^{(2n+1)(n+1)/(n+2)}$
just as in the minimally coupled case \cite{modgrav}, 
and the required accelerated expansion is achieved for $n>2$.
If now $K\equiv K_{\rm DE}$ is taken to be a non-zero constant for large $\varphi$ (that is, $C(\varphi)\approx(K_{\rm DE}/\sqrt 3)\varphi$), then the Jordan scale factor is given by
\begin{equation}
\tilde a\propto \tilde t^{2\over3(1+w_{\rm DE})}\quad
\hbox{with}\quad
w_{\rm DE}=-1+{2\over3}\left({n+2\over n+1}\right)
\left({1+K_{\rm DE}(n+1)\over2n+1+K_{\rm DE}(n+2)}\right).
\label{eqno9}
\end{equation}
Thus there appears another possibility, $K_{\rm DE}\approx -1/(n+1)$, to have late time accelerated expansion besides $K_{\rm DE}\approx0$, $n$ large.

Of course, for the cosmological scenario to be viable, this late time accelerated expansion must be preceded by a matter dominated era with scale factor $\tilde a\propto\tilde t^{2/3}$.
Suppose first that at this matter dominated stage, the function $f(s)$ is such that we can ignore $V(\varphi)$ in (\ref{eqno8}).
This corresponds to the so-called ``$\varphi$-MDE'' phase \cite{Amendola}.
We also assume that the function $K\equiv K_{\rm M}$ is approximately a constant (that is, $C(\varphi)\approx(K_{\rm M}/\sqrt 3)\varphi$).
Setting $a^*\propto t^q$ and $\varphi\sim p\ln t$, one finds from (\ref{eqno8}) that $q=6/((1-K_{\rm M})^2+9)$, $p=2\sqrt{3}(1-K_{\rm M})/((1-K_{\rm M})^2+9)$, and
\begin{equation}
\tilde a\propto {\tilde t}^{\frac{2}{3(1+w_{\rm M})}}\quad
\hbox{with}\quad
w_{\rm M}={2(1-K_{\rm M})^2\over3(3-(1-K_{\rm M})^2)}\,.
\label{eqno10}
\end{equation}

Thus, in the ``tuned'' case when $K_{\rm M}=0$, we obtain the behaviour $\tilde a\propto \tilde t^{1/2}$ during the matter era, which is unacceptable \cite{Amendola}.
The same disqualifying result holds for the detuned case if the matter era takes place when the potential has already reached its asymptotic regime, that is when $\bar s$ is small so that $f(s)\approx -1/(\ell^2\bar s^n)$ and hence $\varphi$ is already large, since the condition for late acceleration is $K\approx0$ (or $K\approx-1/(n+1)$) in that regime.
Therefore, in scenarios when the matter era takes place when $\varphi$ has already reached its asymptotic regime \cite{Amendola2}, detuning is of no help.

However, if the matter era takes place when $\varphi$ is small (this should be realized by an appropriate form of $f(s)$ at large $\bar s$ ($=s\ell^2$)) then we may easily devise a function $K(\varphi)$ which evolves from $K_{\rm M}\approx 1$ for large $\varphi$ to $K_{\rm M}\approx 0$ for small $\varphi$.
In that case we have $\tilde a\propto\tilde t^{2/3}$ during the matter phase.
More specifically, let us take
\begin{equation}
C(\varphi)={\varphi\over\sqrt{3}} 
-{\beta\varphi^2\over2\sqrt{3}}+{\cal O}(\varphi^3)\,,
\label{Cexpand}
\end{equation}
so that $K=1-\beta\varphi+\cdots$ at $\varphi\ll1$.
At leading order, we have $p=0$, hence 
$\varphi$ is constant, $\varphi=\varphi_{\rm M}$ ($\ll1$),
and the Jordan and Einstein frames
 coincide ($\tilde a\approx a^*$ and $\tilde t\approx t$).
Then $p$ to first order is given by
\begin{equation}
p=\frac{2\beta\varphi_{\rm M}}{3\sqrt{3}}\,.
\label{MDEpower}
\end{equation}
It may be noted that this result is perfectly consistent with
Eq.~(\ref{eqno8}).
To the accuracy of our interest, we have
$q=2/3$, hence
\begin{equation}
3H^{*2}={\kappa\rho^*}=\frac{4}{3t^2}\,,\quad
\ddot\varphi+3H^*\dot\varphi=
{\kappa\rho^*}{\beta\varphi_{\rm M}\over2\sqrt{3}}
=\frac{2\beta\varphi_{\rm M}}{3\sqrt{3}t^2}\,.
\end{equation}
Setting $\dot\varphi=p/t$, the second equation
gives $p$ which is consistent with (\ref{MDEpower}).

We note that the only thing we have shown so far is 
the existence of two eras, the ``right'' matter dominated stage
and the accelerated stage at late times.
What remains to be done is to study in detail the dynamical
system (\ref{eqno8}) to see which classes of functions
$f(s)$ and $C(s)$ properly connect the matter era to the late
accelerated era.
This is left for future studies.

\section{Local Gravity Constraints}
The standard way to proceed is first to choose a background solution of the equations of motion, (\ref{KGeq}) and (\ref{Eeq}), then linearise and solve for the perturbations and finally check that the linear approximation was valid \cite{scaten,Chiba}.
It is now well established that if we take
the accelerating cosmological solution of the model
$f(R)=R-\mu^{2(n+1)}/R^n$ \cite{modgrav} as the background,
then the Eddington parameter is $\gamma=1/2$ \cite{Chiba}, 
in gross violation with the solar system 
observations of the Shapiro time delay where it was
found that $\gamma=1$ to better than $10^{-5}$ \cite{Bertotti:2003rm},
that is $\omega>40,000$.
If we follow this standard procedure,
the same result would be obtained also in the detuned case,
since the cosmological solution is the same as in \cite{modgrav}.

Now, as argued in \cite{chamel}, the relevant background when
studying the solar system is not the cosmological solution
but the solution of (\ref{KGeq}) corresponding
to the galactic environment.
Consider a static, spatially homogeneous solution
 $\varphi=\varphi_\mathrm g$ of the Klein--Gordon equation
in a background of a uniform density $\tilde\rho=\tilde\rho_\mathrm g$.
Then we have
\begin{equation}
{\mathrm dV\over\mathrm d\varphi}(\varphi_\mathrm g)
-{1-K(\varphi_\mathrm g)\over2\sqrt{3}}\kappa\rho^*_\mathrm g=0\quad
\hbox{with}\quad
\rho^*={\mathrm e^{4C}\over f'^2}\tilde\rho\,.
\label{galactic}
\end{equation}
Suppose (\ref{galactic}) has a solution for $\varphi_\mathrm g$ small 
 and $K(\varphi_\mathrm g)$ close to unity on galactic scale.
For example, this can be realized by $C(\varphi)$ given by (\ref{Cexpand}).
In this case (\ref{galactic}) gives
\begin{equation}
\varphi_\mathrm g
=\frac{2\sqrt{3}}{\beta}\frac{\mathrm dV/\mathrm d\varphi|_\mathrm g}{\kappa\rho^*_\mathrm g}\,.
\end{equation}
Thus unless there is an extra length scale in the potential $V$
other than the present Hubble scale, the right hand side is
very small, typically of order
$\tilde\rho_\mathrm c/\tilde\rho_\mathrm g=\mathcal O(10^{-5})$ where $\tilde\rho_\mathrm c$ 
is the cosmological density.
As for the Einstein equation (\ref{Eeq}),
its solution on solar scale is almost flat.
Thus we may assume the flat background.

We now linearise the Klein--Gordon equation (\ref{KGeq}) 
around the background solution,
 that is we set $\varphi=\varphi_\mathrm g+\varphi_1$ and 
 $g^*_{ij}=f'_\mathrm g\mathrm e^{-2C_\mathrm g}(\eta_{ij}+h_{ij})$.
We obtain \cite{scaten,Chiba}
\begin{equation}
(\triangle-m^2)\varphi_1
\approx-{4\pi G_{\rm eff}(1-K_\mathrm g)\over\sqrt{3}}\tilde\rho_\odot\quad
\hbox{with}\quad
G_{\rm eff}={\kappa\over8\pi}{\mathrm e^{2C_\mathrm g}\over f'_\mathrm g}\quad
\hbox{and}\quad
m^2=\left.{f'_\mathrm g\over\mathrm e^{2C_\mathrm g}}{\mathrm d^2V\over\mathrm d\varphi^2}\right|_\mathrm g\,,
\label{eqno12}
\end{equation}
where $\tilde\rho_\odot$ is the density of the Sun
and $\triangle$ the standard $3$-Laplacian.
Typically (see the example below) $m^2={\cal O}(\kappa\tilde\rho_\mathrm g)$ or smaller.
Since $\triangle={\cal O}(L^{-2})\gg m^2$, where $L$ is the typical solar system scale, the scalaron mass $m$ is negligible.

We emphasise that this is where our ``chameleon detuning'' differs from the standard one.
Indeed, in the standard $f(R)$ case 
the Brans--Dicke function is zero (or too small as in \cite{Tsujikawa}) 
and one has to invoke, on top of the galactic environment,
 a complicated nonlinear effect inside the Sun.
Here, to the contrary, the Brans--Dicke function $\omega$ is large
and the mass $m$ is small so that we remain within the regime of
the linear approximation.
 
 For negligible $m^2$, the solution of (\ref{eqno12}) is 
$\varphi_1\approx {G_{\rm eff}(1-K_\mathrm g)}{M_\odot/(\sqrt{3}r)}$.
 Since $\varphi_1\ll\varphi_\mathrm g$ we check that we could use the 
linear approximation.
 
 As for the Einstein equation (\ref{Eeq}), the linearised
equation $\delta G^*_{ij}\approx\kappa T^*_{{\odot}ij}$ 
gives the (linearised) Schwarzschild metric
 (see \cite{scaten} for a detailed calculation),
\begin{equation}
\mathrm ds^2_*=
\frac{f'_\mathrm g}{\mathrm e^{2C_\mathrm g}}\left[
 -\left(1-\frac{2G_{\rm eff}M_\odot}{r}\right)\mathrm dt^2
+\left(1+\frac{2G_{\rm eff}M_\odot}{r}\right)\mathrm d\vec x^2\right],
\end{equation}
yielding the Jordan frame metric:
\begin{equation}
\mathrm d\tilde s^2=\frac{\mathrm e^{2C}}{f'}\mathrm ds_*^2
=-\left(1-{2\tilde G M_\odot\over r}\right)\mathrm dt^2
+\left(1+{2\gamma\tilde G M_\odot\over r}\right)\mathrm d\vec{x}^2\,,
\end{equation}
with $\tilde G=G_{\rm eff}(1+(1-K_\mathrm g)^2/3)$ and
\begin{equation}
\gamma-1=-{2(1-K_\mathrm g)^2\over 3+(1-K_\mathrm g)^2}\,.
\label{eqno13}
\end{equation}
Hence for standard, minimally coupled $f(R)$ models 
with $K=0$, we recover $\gamma=1/2$ \cite{Chiba}.
We also see that, in order to comply with 
the Cassini mission result \cite{Bertotti:2003rm},
we must have $|1-K_\mathrm g|<0.01$.

Using ``detuning'' we have therefore found a way to evade local gravity constraints, using one aspect of the ``chameleon'' mechanism \cite{chamel,Amendola2}, that is the fact that the solar system galactic environment is much denser than the cosmological background, while remaining within the well understood linear approximation.

In fact, this ``chameleon-like'' detuning may even help to reconcile 
the most emblematic model of $f(R)$ dark energy \cite{modgrav},
\begin{equation}
f(s)={1\over\ell^2}\left(\bar s-{1\over \bar s^n}\right)\quad
\hbox{so that}\quad
V(\varphi)={n+1\over4\ell^2n^{n\over n+1}}
\mathrm e^{-4\varphi/\sqrt{3}}(\mathrm e^{2\varphi/\sqrt{3}}-1)^{n\over n+1}\,,
\label{eqno14}
\end{equation}
with local gravity constraints.
For the sake of the example, let us consider the following coupling function:
\begin{equation}
C(s)={ f'(s)-1\over 2f'(s)}\quad
\hbox{that is}\quad
K(s)={1\over f'(s)}\quad
\hbox{with}\quad
f'=\mathrm e^{2\varphi/\sqrt{3}}\,.
\label{eqno15}
\end{equation}
As can easily be seen, we have then that $\rho^*\approx\tilde\rho$ and $\varphi_\mathrm g\approx{\sqrt{3}/(2\ell^2\kappa\tilde\rho_\mathrm g)}={\cal O}(\tilde\rho_\mathrm c/\tilde\rho_\mathrm g)$ for large $n$.
Hence $m^2\approx{\mathrm d^2V/\mathrm d\varphi^2|_\mathrm g}\approx
 -(2n\sqrt{3}\ell^2\varphi_\mathrm g)^{-1}={\cal O}(\kappa\tilde\rho_\mathrm g/n)$
($m^2<0$ but is small, yielding a Dolgov--Kawasaki 
instability \cite{DolKawa} setting in on a galactic time scale).
Finally $1-K_\mathrm g\approx 2\varphi_\mathrm g/\sqrt{3}$ so that
 $\gamma-1\approx-8\varphi_\mathrm g^2/9={\cal O}(10^{-10})$
 well below the Cassini bound.

\section{Conclusion}
We have introduced a chameleon-like coupling of matter to the scalaron degree of freedom of $f(R)$ gravity and argued that such non-minimal coupling might retain the late cosmological acceleration they predict while rendering them compatible with local gravity constraints.
Of course many aspects of these ``detuned'' $f(R)$ gravity theories remain to be explored, starting with a detailed analysis of the dynamical system (\ref{eqno8}) in the line of \cite{Amendola2} and its cosmological perturbations,  using $f(R)$ functions such as those proposed in \cite{HuSaStaron}.

While we were writing up this paper, we became aware of \cite{Tsujikawa} where, in contrast with the present proposal, the coupling function $K$ (called $Q$ in \cite{Tsujikawa}) is a given constant.

\begin{acknowledgments}
ND thanks the Racah Institute of Physics and the Lady Davis Foundation
for their warm hospitality and generous support when this work was performed.
MS was supported in part by JSPS Grant-in-Aid for
Scientific Research (B) No.~17340075, (A) No.~18204024, 
and by JSPS Grant-in-Aid for Creative Scientific Research No.~19GS0219.
YS was supported by Grant-in-Aid for JSPS Fellows No.~19-7852.
This work was also supported in part by a CNRS--JSPS collaboration
program.

We are grateful to Alexei Starobinsky for comments on the first version of this paper.
\end{acknowledgments}

\end{document}